\documentclass[prd, preprint, 11pt, nofootnoteinbib]{revtex4-1}
\usepackage{hyperref}
\usepackage{amsmath}
\usepackage{amsfonts}
\usepackage[dvips]{graphicx}
\usepackage{mdframed}
\usepackage{tensor}
\usepackage[retainorgcmds]{IEEEtrantools}
\usepackage{amssymb}

\begin{document}

\begin{center}
 { \large {\bf Stochastic modification of the Schr\"odinger-Newton equation}}


\vskip 0.3 in

{\large{\bf Sayantani Bera$^*$, Ravi Mohan$^{*,a}$ and Tejinder P.  Singh}}

{\it Tata Institute of Fundamental Research,}
{\it Homi Bhabha Road, Mumbai 400005, India}\\
$^a$Address after September 1, 2015:  {\it University of Texas at Austin, Austin, TX 78712, USA}

* Equal contribution by both authors

\bigskip
{\tt sayantani.bera@tifr.res.in, ravimohan1991@gmail.com, tpsingh@tifr.res.in}\\

\end{center}

\bigskip
\bigskip

\centerline{\bf ABSTRACT}
\noindent The Schr\"{o}dinger-Newton [SN] equation describes the effect of self-gravity on the evolution of a quantum system, and it has been proposed that gravitationally induced decoherence drives the system to one of the stationary solutions of the SN equation. However, the equation by itself lacks a decoherence mechanism, because it does not possess any stochastic feature. In the present work we derive a stochastic modification of the Schr\"{o}dinger-Newton equation, starting from the Einstein-Langevin equation in the theory of stochastic semiclassical  gravity. We specialize this equation  to the case of a single massive point particle, and by using Karolyhazy's phase variance method, we derive the Di\'osi - Penrose criterion for the decoherence time. We obtain a (nonlinear) master equation corresponding to this stochastic SN equation. This equation is however linear at the level of the approximation we use to prove decoherence; hence the no-signalling requirement is met. Lastly, we use physical arguments to obtain expressions for the decoherence length of extended objects.  

\noindent 

\section{Introduction}
A problem of long-standing interest has been to understand if gravity can help solve the quantum measurement problem; or even if it does not actually cause collapse of the wave-function, can it at least be a source of decoherence. The physical picture being that the gravitational field of space-time cannot be exactly classical but must possess intrinsic quantum uncertainty. Can this uncertainty relate to the quantum evolution described by the Schr\"{o}dinger equation in such a way that for macroscopic masses the space-time uncertainty induces decoherence and position localisation? The answer seems to be in the affirmative, as has been demonstrated in various studies by Karolyhazy \cite{Karolyhazi:66} and then followed by Karolyhazy and collaborators in several papers \cite{Karolyhazi:86, Karolyhazy:74, Karolyhazy:90, Karolyhazy:95, Karolyhazy:1982, Frenkel:77, Frenkel:90, Frenkel:95, Frenkel:2002, Frenkel:97},  and also shown by Di\'osi and collaborators  \cite{Diosi:87, Diosi:07, Diosi:89, Diosi:87a}. These authors demonstrate, subject to crucial assumptions about how quantum matter affects gravity, that there is an inherent uncertainty in the space-time geometry, whose origin lies in the quantum nature of the sources which produce them. This uncertainty is modelled by introducing a classical stochastic potential in the Schr\"{o}dinger equation, and it is shown that the stochasticity induces decoherence, whose properties depend on the mass and size of the quantum object under study. The work of Karolyhazy and of Di\'osi has recently been compared in \cite{Bera:2014}.

Having considered  the effect of the space-time uncertainty produced by other objects, on the Schr\"{o}dinger equation, it is also important to consider  the effect of self-gravity of a quantum object on its Schr\"{o}dinger evolution. This self-gravity has an intrinsic quantum uncertainty and although we do not quite know at present how to describe the self-gravity of a quantum object, one can attempt to model it.  One well-known approach is the  Schr\"{o}dinger-Newton [SN] equation. It is proposed  that the self-gravitational potential $V$ produced by a quantum source in state $\Psi$ satisfies a semiclassical Poisson equation
\begin{equation}
\nabla^{2}V = 4\pi G m |\Psi|^2
\label{potnal}
\end{equation} 
whose solution is incorporated  in  the potential dependent Schr\"{o}dinger equation
\begin{equation}
i\hbar\frac{\partial\Psi}{\partial t} = - \frac{\hbar^2}{2m} \nabla^{2}\Psi + mV \Psi
\label{sceqn}
\end{equation}
This gives the Schr\"{o}dinger-Newton equation \cite{Penrose:96, Penrose:98, Penrose:00, Diosi:84}
\begin{equation}
i\hbar \frac{\partial \Psi({\bf r}, t)}{\partial t} =  - \frac{\hbar^2}{2m} \nabla^{2}\Psi ({\bf r}, t) -
Gm^2 \int \frac{ |\Psi ({\bf r'},t)|^2}{|{\bf r}-{\bf r'}|} d{\bf r'} \Psi ({\bf r},t)
\label{sneqn}
\end{equation}
This equation has been discussed extensively in the literature \cite{Bernstein:98,giulini2011gravitationally,Harrison:2003,Moroz:98,Ruffini:69,giulini:2012,giulini:2013,Hu:2014,Anastapoulos:2014, Bahrami:2014, Colin:2014}.
The equation is nonlinear and deterministic. This makes it different in character from the Karolyhazy equation and the Di\'osi equation, where by virtue of the stochastic potential the models predict decoherence at the level of the master equation. The SN equation does not predict decoherence but is instead suggested as an equation whose stationary solutions are  the ones to which the system will proceed upon decoherence, once a gravity-based decoherence mechanism has been incorporated in this system. What has been shown is that there is a gravitationally induced inhibition of dispersion of an expanding wave-packet, at the critical length $a_c \sim \hbar^2 / Gm^3$ \cite{giulini2011gravitationally}. Another concern which can be raised about the SN equation is that its deterministic non-linear structure allows for superluminal signalling.  For a recent critique of the SN equation see 
\cite{Anastapoulos:2014} and  \cite{Bahrami:2014}.

The reason why decoherence is not observed  in the SN equation is evident: one is only taking into account the mean potential in the semiclassical Poisson equation, whereas the Karolyhazy and Di\'osi models incorporate stochastic fluctuations by way of the stochastic potential in the Schr\"{o}dinger equation, which is modelled after the space-time geodesic uncertainty.  We need to take into consideration  the stochastic fluctuations, apart from the mean, and modify the SN equation accordingly. One possible way to do this is to start from the theory of stochastic gravity \cite{Hu:2004}, which takes into account corrections to the semiclassical Einstein equations
\begin{equation}
R_{\mu\nu} -\frac{1}{2} g_{\mu\nu} R = \frac{8\pi G}{c^4} \langle\Psi| \hat{T}_{\mu\nu}|\Psi\rangle
\label{semieinstein}
\end{equation}
by considering the role of the two-point fluctuations of the energy-momentum tensor [going from semiclassical gravity to the Einstein-Langevin equation]. The purpose of the present paper is to develop this idea and propose a new stochastic SN equation. The effect of quantum fluctuations of the energy-momentum tensor on the geometry is modelled by defining a classical stochastic field. This stochastic field eventually acts as a source for the gravitational potential (which is now stochastic) via a modified semiclassical Poisson equation. This stochastic potential is included in the Schr\"{o}dinger equation, in the spirit of Karolyhazy's and Di\'osi's work, and as we shall show, it leads to a stochastic SN equation which produces decoherence in the quantum evolution.

The plan of the paper is as follows. In Section II we recall the Einstein-Langevin equation. In Section III we use the Einstein-Langevin equation to derive a new stochastic SN equation. Section IV uses this new equation to provide a proof of gravitational decoherence for the gaussian state of a point particle.  In Section V we obtain elementary estimates for decoherence scales for extended objects while working with the SN equation. Concluding remarks are in Section VI. Appendix I derives the master equation for the density matrix, corresponding to the stochastic SN equation. Appendix II gives details of the decoherence calculation reported in Section IV.

A stochastic modification of the Schr\"{o}dinger equation has also been proposed in \cite{Nimmrichter:2015}. We contrast our work with that of \cite{Nimmrichter:2015}  in the concluding section. The application of stochastic gravity to the measurement problem has recently been studied in \cite{Hu:2015}.
The possible role of gravity in the decoherence / collapse of the wave function has been reviewed, amongst other places, in \cite{RMP:2012, gao, Singh:2015}.

\section{Semiclassical gravity and the Einstein-Langevin equation}
Semiclassical gravity describes the interaction of the gravitational field, which is treated classically, with quantum matter fields. The field equation for the classical metric is the semiclassical Einstein equation, which gives the back reaction of the matter fields on the spacetime; it is a generalization of the Einstein equation where the source is \emph{the expectation value in some quantum state of the matter stress-energy tensor operator}. The coupling of gravitational field to matter is modelled by the semiclassical Einstein equations (\ref{semieinstein}).

We will be interested in the weak-field limit 
\begin{IEEEeqnarray}{rCl}
g_{\mu\nu} = \eta_{\mu\nu} + h_{\mu\nu},\label{weak field}
\end{IEEEeqnarray}
where $|h_{\mu\nu}|\ll 1$ and $\eta_{\mu\nu}$ is the  Minkowski metric; thus we obtain the \emph{linearised theory of gravity}. In a weak field situation one can expand the field equations in powers of $h_{\mu\nu}$ using a coordinate frame where Eqn. (\ref{weak field}) holds and one can retain only linear terms. 
Using Eqns. (\ref{semieinstein}) and (\ref{weak field}), one can arrive at the differential equation for the metric field $h_{\mu\nu}$, given by \cite{Bahrami:2014} 
\begin{IEEEeqnarray}{rCl}
\Box h_{\mu\nu} = -\frac{16\pi G}{c^4}\left(\langle \Psi |\hat{T}_{\mu\nu}|\Psi\rangle - \frac{1}{2}\eta_{\mu\nu}\langle \Psi |\hat{T}_{\alpha\beta}\eta^{\alpha\beta}|\Psi\rangle\right).
\end{IEEEeqnarray}
In the Newtonian limit the $\langle \Psi |\hat{T}_{00}|\Psi\rangle$ component is large compared to other components of the stress-energy tensor. Hence in this limit the field equation for linearised gravity reduces to a semiclassical Poisson equation of the form
\begin{IEEEeqnarray}{rCl}
\nabla^2V=\frac{4\pi G}{c^2}\langle \Psi |\hat{T}_{00}|\Psi\rangle,
\end{IEEEeqnarray}
where $h_{00}{c^2}/{2} \equiv -V$.
Using $\hat{T}_{00} = c^2{\varrho}$ we get the familiar potential field used in the 
Schr\"{o}dinger-Newton equation, 
as explained in \cite{Bahrami:2014}. The same result also follows by noting that for a single particle we have 
${\varrho} = m |{\bf r} \rangle\langle {\bf r} |$. In this manner we can think of the SN equation (\ref{sneqn}) as the non-relativistic limit of semiclassical gravity. 

The semiclassical Einstein equations have as a source only the `mean field' coming from the expectation value of the quantum stress-energy tensor. It has been argued that there can be circumstances where quantum fluctuations of the stress-energy  tensor  about the mean can be important, and then the semiclassical theory should be modified. One possible modification is the so-called theory of stochastic gravity. Here,
the Einstein-Langevin equation \cite{Hu:2004, Hu2:2002, Rosario:99} includes also the back reaction on the space-time metric of the lowest order stress-energy quantum fluctuations. This results in an effective theory which predicts linear stochastic corrections to the semiclassical metric. The equations include a Gaussian stochastic tensor field  $\xi_{\mu\nu}(x)$ representing the stress-energy fluctuations, and having the following three properties:
\begin{enumerate}
\item The stochastic average $\langle.\rangle_{s}$ of the field vanishes i.e.
\begin{IEEEeqnarray}{rCl}
\langle \xi_{\mu\nu}\rangle_{s} = 0.
\end{IEEEeqnarray}

\item The two point correlation of this field is given by
\begin{IEEEeqnarray}{rCl}
\langle \xi_{\alpha\beta}(x)\xi_{\mu\nu}(y)\rangle_{s} = N_{\alpha\beta\mu\nu}(x,y).\label{2pointcorre}
\end{IEEEeqnarray}
Here $x$ and $y$ are space-time four vectors.  $N_{\alpha\beta\mu\nu}(x,y)$ is called the noise kernel. It is related to the stress-energy tensor in the following way
\begin{IEEEeqnarray}{rCl}
8N_{\alpha\beta\mu\nu}(x,y) = \langle\{\hat{t}_{\alpha\beta}(x),\hat{t}_{\mu\nu}(y)\}\rangle,\label{noisekernel}
\end{IEEEeqnarray}
where $\{.\}$ is the anti-commutator, $\langle.\rangle$ is the expectation value and $\hat{t}_{\alpha\beta} = \hat{T}_{\alpha\beta}-\langle\hat{T}_{\alpha\beta}\rangle$.

\item Higher moments and cumulants of this stochastic field vanish.
\end{enumerate}

The metric continues to be {\it classical}, and now becomes a stochastic tensor field, sourced by the mean as well as by the fluctuations. The Einstein-Langevin equation takes the form
\begin{IEEEeqnarray}{rCl}
R_{\mu\nu}-\frac{1}{2}g_{\mu\nu}R = \frac{8\pi G}{c^4}\left(\langle \Psi |\hat{T}_{\mu\nu}|\Psi\rangle + 2\xi_{\mu\nu}\right)
\label{ELeqn} 
\end{IEEEeqnarray}
with the last term on the right hand side incorporating the effect of matter fluctuations. 

We will now see how the inclusion of matter fluctuations modifies the SN equation, leading to our proposal for a new stochastic SN equation.

\section{A proposal for a stochastic Schr\"{o}dinger-Newton equation}
We consider the weak field limit of the Einstein-Langevin equation (\ref{ELeqn}). 
On applying the weak-field condition Eqn. (\ref{weak field}) in (\ref{ELeqn}), we obtain the following field equation for the metric $h_{\mu\nu}$
\begin{IEEEeqnarray}{rCl}
\Box h_{\mu\nu} = -\frac{16\pi G}{c^4}\left(\langle \Psi |\hat{T}_{\mu\nu}|\Psi\rangle - \frac{1}{2}\eta_{\mu\nu}\langle \Psi |\hat{T}_{\alpha\beta}\eta^{\alpha\beta}|\Psi\rangle + 2\xi_{\mu\nu}-\xi_{\alpha\beta}\eta^{\alpha\beta}\eta_{\mu\nu}\right)
\end{IEEEeqnarray}
In the Newtonian limit, we get the equation
\begin{IEEEeqnarray}{rCl}
\nabla^2 V({\bf r}, t) = \frac{4\pi G}{c^2}\left(\langle \Psi |\hat{T}_{00}|\Psi\rangle + 2(\xi_{00}+\frac{1}{2}\xi_{\alpha\beta}\eta^{\alpha\beta})\right)
\end{IEEEeqnarray}
Now if we consider only the $\xi_{00}$ component (since Newtonian limit has been assumed), we get
\begin{IEEEeqnarray}{rCl}
\nabla^2 V({\bf r}, t) = \frac{4\pi G}{c^2}\left(\langle \Psi |\hat{T}_{00}|\Psi\rangle + \xi_{00}\right).
\end{IEEEeqnarray}
As expected, we now have a stochastic source contributing towards the gravitational potential. Using Green's function, we derive the modified potential which includes a stochastic component  \begin{IEEEeqnarray}{rCl}
V({\bf r},t) = -Gm\int\frac{|\Psi({\bf r}^{\,\prime},t)|^2}{|{\bf r}-{\bf r}^{\,\prime}|}d{\bf r}^{\,\prime} - \frac{G}{c^2}\int\frac{\xi_{00}({\bf r}^{\,\prime},t)}{|{\bf r}-{\bf r}^{\,\prime}|}d{\bf r}^{\,\prime}.\label{stochasticeq}
\end{IEEEeqnarray}
Using this potential in the Schr\"{o}dinger equation (\ref{sceqn}) gives the new stochastic Schr\"{o}dinger-Newton equation
\begin{IEEEeqnarray}{rCl}
\label{newstsn}
i\hbar\frac{\partial}{\partial t}\Psi({\bf r},t) = -\frac{\hbar^2}{2m}\nabla^2\Psi({\bf r},t)- Gm^2\int\frac{|\Psi({\bf r}^{\,\prime},t)|^2}{|{\bf r}-{\bf r}^{\,\prime}|}d{\bf r}^{\,\prime}\Psi({\bf r},t) - \frac{Gm}{c^2}\int\frac{\xi_{00}({\bf r}^{\,\prime},t)}{|{\bf r}-{\bf r}^{\,\prime}|}d{\bf r}^{\,\prime}\Psi({\bf r},t)\
\end{IEEEeqnarray}
This equation can also be  worked out by going from field theory to quantum mechanics of $N$ particles in non-relativistic limit and then obtaining the SN equation for single particle, following the method outlined in \cite{Bahrami:2014}. The modified potential, where the modifying stochastic component originates in the fluctuations of the stress tensor, is responsible for the stochastic SN equation.

A couple of striking similarities with Diosi's model and Karolyhazy's model can be immediately recognised. First, the total stochastic potential $V({\bf r}, t)$ is a linear sum of the Newtonian potential and stochastic noise (which is Gaussian).
Second, the average of the stochastic component of the potential vanishes due to the first property of $\xi_{\mu\nu}(x)$. In a sense then, the stochastic SN equation (unlike the SN equation itself) is on the same footing as the stochastic Schr\"{o}dinger equations of the Karolyhazy model and the Di\'osi model. The key difference between the first and the latter two is that while one accounts for effects of self-gravity, the other two account for stochastic fluctuations induced by extrinsic space-time uncertainty. To our understanding, a more general treatment should take into account effects of both self-gravity and extrinsic uncertainty in the same scheme. We hope to pursue such an investigation in future work.

For use in the next section, we work out the two point correlation of the stochastic potential. 
Let the stochastic component of the gravitational potential in Eqn. (\ref{stochasticeq}) be denoted by$V_{st}({\bf r}, t)$ and written as
\begin{IEEEeqnarray}{rCl}
V_{st}({\bf r},t) = \frac{G}{c^2}\int\frac{\xi_{00}({\bf r}^{\,\prime},t)}{|{\bf r}-{\bf r}^{\,\prime}|}d{\bf r}^{\,\prime},
\end{IEEEeqnarray}
Thus the correlation takes the form
\begin{IEEEeqnarray}{rCl}
\langle V_{st}({\bf r},t)V_{st}({\bf r}^{\,\prime},t)\rangle_{s} = \frac{G^2}{c^4}\int\frac{\left\langle\xi_{00}({\bf x},t)\xi_{00}({\bf x}^{\,\prime},t)\right\rangle_{s}}{|{\bf x}-{\bf r}||{\bf x}^{\,\prime}-{\bf r}^{\,\prime}|}d{\bf x}^{\,\prime}d{\bf x}.
\end{IEEEeqnarray}
Using Eqn. (\ref{2pointcorre}) for the definition of $\xi_{00}$
\begin{IEEEeqnarray}{rCl}
\langle \xi_{00}(x)\xi_{00}(y)\rangle_{s} = N_{0000}(x,y),
\end{IEEEeqnarray}
and Eqn. (\ref{noisekernel})
\begin{IEEEeqnarray}{rCl}
N_{0000}(x,y) = \frac{1}{8}\langle\{\hat{t}_{00}(x),\hat{t}_{00}(y)\}\rangle,\label{2pointcorrekernel}
\end{IEEEeqnarray}
where $\{.\}$ is the anti-commutator, $\langle.\rangle$ is the expectation value and $\hat{t}_{00} = \hat{T}_{00}-\langle\hat{T}_{00}\rangle$, we arrive at
\begin{IEEEeqnarray}{rCl}
\langle V_{st}({\bf r},t)V_{st}({\bf r}^{\,\prime},t)\rangle_{s} = \frac{G^2}{8c^4}\int\frac{\langle\Psi|\{\hat{T}_{00}-\langle\hat{T}_{00}\rangle,\hat{T}_{00}-\langle\hat{T}_{00}\rangle\}|\Psi\rangle}{|{\bf x}-{\bf r}||{\bf x}^{\,\prime}-{\bf r}^{\,\prime}|}d{\bf x}^{\,\prime}d{\bf x}.
\label{fullcorr}
\end{IEEEeqnarray}

Using the form $\hat{T}_{00}({\bf x}) = c^2\hat{\varrho}=mc^2|{\bf x}\rangle\langle{\bf x}|$ for a single particle, we get
\begin{IEEEeqnarray}{rCl}
\langle V_{st}({\bf r},t)V_{st}({\bf r}^{\,\prime},t)\rangle_{s} = \frac{G^2m^2}{8}\bigg[2\int\frac{|\psi({\bf x},t)|^2}{|{\bf x}-{\bf r}||{\bf x}-{\bf r}^{\,\prime}|}d{\bf x}\: -2\int\frac{|\psi({\bf x},t)|^2|\psi({\bf x}^{\,\prime},t)|^2}{|{\bf x}-{\bf r}||{\bf x}^{\,\prime}-{\bf r}^{\,\prime}|}d{\bf x}^{\,\prime}d{\bf x}\bigg].\label{correlationpot}
\end{IEEEeqnarray}

We can work out the master equation for the density matrix, following the methods of \cite{Adler:07}.
The  evolution of the state vector in (\ref{newstsn}) can be written as follows
\begin{equation}
\frac{d|\Psi(t)\rangle}{dt} =\left[-\frac{i}{\hbar}\hat{H}_0 +i\sqrt{\gamma}\int d{\bf k}\langle \hat{A}^\dagger({\bf k})\rangle\hat{A}({\bf k}) + i\sqrt{\gamma}\int d{\bf k} \tilde{\xi}({\bf k},t)\hat{A}({\bf k})\right] |\Psi(t)\rangle,\label{stochasticsnvector}
\end{equation}
where $\hat{A}({\bf k}) = (m/k)e^{i{\bf k}.\hat{{\bf r}}}$, $\sqrt{\gamma}=(G/2\pi^2\hbar)$, $\langle\hat{A}^\dagger({\bf k})\rangle = \langle\Psi|\hat{A}^\dagger({\bf k})|\Psi\rangle$ and where
\begin{equation}
\tilde{\xi}({\bf k},t) = \frac{k}{c^2}\int d{\bf r}e^{-i{\bf k}.{\bf r}}\int d{\bf r}^{\,\prime} \frac{\xi_{00}({\bf r}^{\,\prime},t)}{|{\bf r}-{\bf r}^{\,\prime}|},
\end{equation}
is the non-white stochastic noise. Using the techniques of \cite{Adler:07}, the non-Markovian master equation for the evolution of the density matrix corresponding to Eqn. (\ref{stochasticsnvector}) can be shown to be
\begin{equation}
\begin{split}
\frac{d\hat{\rho}}{dt} &= -\frac{i}{\hbar}[\hat{H},\hat{\rho}] +i\sqrt{\gamma}\int d{\bf k}\left(\langle\hat{A}^\dagger\rangle\hat{A}\hat{\rho}-\hat{\rho}\hat{A}^\dagger\langle\hat{A}\rangle\right)\\
&-\gamma\int_0^t dt^\prime d{\bf k}d{\bf k}^{\,\prime} D({\bf k},{\bf k}^{\,\prime},t,t^\prime)\bigg(\hat{A}({\bf k})\hat{A}({\bf k}^{\,\prime},t^\prime-t)\hat{\rho}+\hat{\rho}\hat{A}^\dagger({\bf k}^{\,\prime},t^\prime-t)\hat{A}^\dagger({\bf k})\bigg),
\end{split}
\end{equation}
where $\hat{\rho} = \langle|\Psi\rangle\langle\Psi|\rangle_s$. $\hat{A}({\bf k}^{\,\prime},t^\prime-t)$ 
is the interaction picture operator evolved up to time $(t'-t)$.
The proof  for the derivation of this master equation is given in Appendix I. 

%

It is challenging to demonstrate decoherence from the above master equation, but an alternative method applied by Karolyhazy, which we call the phase variance method, comes to our rescue. The equivalence of the phase variance method and the Markovian master equation of Di\'osi's model has been demonstrated by us in \cite{Bera:2014} and it is expected that this equivalence holds in general for non-Markovian master equations as well.

\section{Phase variance method and proof of decoherence}
In this section we adopt the scheme used by Karolyhazy in his work, to find the decoherence effect due to the stochastic potential in the stochastic SN equation (\ref{newstsn}).
In our analysis, we have found that
\begin{equation}
i\hbar \frac{\partial\Psi({\bf r},t)}{\partial t} = -\frac{\hbar^2}{2m}\nabla^2 \Psi({\bf r},t) + V({\bf r},t)\Psi({\bf r},t)
\label{ssn}
\end{equation}
where 
\begin{equation}
V({\bf r},t)= -Gm^2 \int \frac{|\Psi({\bf r}',t)|^2}{|{\bf r}-{\bf r}'|} d\,{\bf r'} - \frac{Gm}{c^2}\int \frac{\xi_{00}({\bf r}',t)}{|{\bf r}-{\bf r}'|} d\,{\bf r'}
\end{equation}

We now make an important approximation, in order to be able to make progress. In the potential $V$, the exact state $\Psi$ will be replaced by the free wave function, denoted as  $\psi$, this being the solution of the Schr\"{o}dinger equation with gravitational back-reaction ignored (i.e. $V=0$). Thus the gravitational effect is being calculated iteratively. In other words, in our further calculations we will use
\begin{equation}
V({\bf r},t)= -Gm^2 \int \frac{|\psi({\bf r}',t)|^2}{|{\bf r}-{\bf r}'|} d\,{\bf r'} - \frac{Gm}{c^2}\int \frac{\xi_{00}({\bf r}',t)}{|{\bf r}-{\bf r}'|} d\,{\bf r'}
\label{vtot}
\end{equation}
This has the important consequence that the stochastic Schr\"{o}dinger-Newton equation becomes linear, as is the case for the models of Karolyhazy and Di\'{o}si. Thus the deterministic nonlinear SN equation has now been replaced by a stochastic linear SN equation.  The corresponding master equation for the density matrix also becomes linear in the density matrix, in this approximation. To our understanding, having a linear master equation accompanying a stochastic linear SN equation alleviates the problem of superluminal signalling faced by the original deterministic nonlinear SN equation.

Following the argument of Karolyhazy, we can demonstrate that the solution of this linear stochastic equation can be given in the form 
\begin{equation}
\Psi({\bf r},t) = \psi({\bf r},t) \exp (i\phi_{st}({\bf r},t))
\label{adans}
\end{equation}
where $\psi({\bf r},t)$ is the free solution without any gravitational back-reaction and $\phi_{st}({\bf r},t)$ is the phase of the actual solution with respect to the free wavefunction, and is given by,
\begin{equation}
\phi_{st}({\bf r},t) = -\frac{1}{\hbar} \int^t_0 V({\bf r},t') d\,t'
\label{phaseap}
\end{equation}

For completeness, we outline the justification of this result of Karolyhazy. The Schr\"{o}dinger equation in the presence of the stochastic potential is given by
\begin{equation}
i\hbar\frac{\partial \Psi ({\bf r},t)}{\partial t} = H({\bf r},t)\Psi({\bf r},t)
\end{equation} 
where the time dependent Hamiltonian is given by,
\begin{equation}
H({\bf r},t) = H_0({\bf r}) + V({\bf r},t)
\end{equation}
where the second term denotes a small perturbation around the time-independent Hamiltonian.
To solve this equation, Karolyhazy considers the perturbation to be switched on adiabatically (see, for instance, Karolyhazy's detailed paper of 1990, pg. 223 \cite{Karolyhazy:90}). Hence we can assume the ``adiabatic approximation" to be valid i.e., if a system starts with a certain state, it will remain in that same state after the perturbation is introduced adiabatically.

Now, let the initial state be $\Psi({\bf r},0) = \psi ({\bf r},0)$ where $\psi({\bf r},t)$ is the solution without the perturbation so that $\psi ({\bf r},t) = \psi({\bf r},0) \exp (-iEt/\hbar)$. If now, the perturbation is turned on adiabatically, the system essentially remains in the same state, with only some phase factors introduced, similar to the unperturbed case.
Using the adiabatic approximation, we can write (see Griffiths \cite{Griffiths:2014} Chap. 10, Eqns. 10.12, 10.13 and 10.23)
\begin{equation}
\Psi({\bf r},t)= \psi({\bf r},0) \exp[i\theta({\bf r},t)] \exp[i\gamma({\bf r},t)]
\end{equation}
where $\theta({\bf r},t) = -\frac{1}{\hbar}\Big[E_0 t + \int_0^t V ({\bf r},t') \,dt' \Big]$, $E_0$ being the unperturbed eigenenergy and $V$ is due to perturbation. The phase $\gamma({\bf r},t)$ is given by Eqn 10.22 of Griffiths \cite{Griffiths:2014}.
Hence the above equation can be rewritten as
\begin{equation}
\Psi({\bf r},t)= \psi({\bf r},t) \exp\Big[-\frac{i}{\hbar}\int_0^t V ({\bf r},t') \,dt'\Big] \exp[i\gamma({\bf r},t)]
\end{equation}
The last phase term $\gamma({\bf r},t)$ (called the geometric phase) can be neglected as it depends weakly on ${\bf r}$ (see Karolyhazy \cite{Karolyhazy:90}). Thus we recover the desired solution written above in Eqns. (\ref{adans}) and (\ref{phaseap}). 

A few remarks are in order, with regard to the application of this phase variance method to the present case of the stochastic SN equation (\ref{ssn}). Both the terms in the potential (\ref{vtot}), namely the mean as well as the fluctuation, are treated on the same footing \cite{Hu:2004}, as in principle there can be circumstances when they are comparable. Thus, from the viewpoint of application of the adiabatic approximation assumed while writing down the ansatz (\ref{adans}) it is assumed that the total potential can be treated as a perturbation on the free particle. While this is definitely evident for microscopic objects, the assumption maybe considered to be a reasonable one in the macroscopic case as well, so long as the gravitational field in question is sufficiently weak. The plausibility of the results that we derive below serves to justify the validity of the ansatz (\ref{adans}). 

The significance of the form (\ref{adans}) of the solution to the stochastic SN equation is the following. The phase $\phi_{st}({\bf r}, t)$ at any spatial point ${\bf r}$ is a stochastic variable. If we consider two fixed spatial points ${\bf r_1}$ and ${\bf r_2}$, the phase difference between the quantum states at these two points is also stochastic, and the variance of this difference grows with time. When the variance becomes of the order $\pi^2$, we say that the states at these two points decohere. This gives the decoherence time as a function of the spatial distance $|{\bf r_1}-{\bf r_2}|$. This decoherence criterion has been  shown  to be equivalent \cite{Bera:2014} to the one defined using a Markovian master equation, as in Di\'osi's model, and we assume that the equivalence holds  in the present non-Markovian case as well. The plausibility of the assumption is supported by the nature of the results we find below, which tally with results obtained by others earlier.

We now calculate the phase variance between two points ${\bf r}_1$ and ${\bf r}_2$ with the formula
\begin{equation}
\Delta \phi^2 = \langle[\phi_{st}({\bf r_1},t) - \phi_{st}({\bf r_2},t)]^2\rangle_s
\end{equation}
We find the time $t$ for which the above quantity would be $\sim \pi^2$ and that will give us the decoherence time.
Calculating the variance using the definitions (\ref{phaseap}) and (\ref{vtot}) we get,
\begin{equation}
\begin{split}
\Delta \phi^2 &= \frac{3G^2m^4}{4\hbar^2}\Big[\int\int \frac{|\psi({\bf r}',t')|^2 |\psi({\bf r}'',t'')|^2}{|{\bf r_1}-{\bf r}'| |{\bf r_1}-{\bf r}''|} d\,^3 r' d\,^3 r'' d\,t' \,dt'' \\
& + \int\int \frac{|\psi({\bf r}',t')|^2 |\psi({\bf r}'',t'')|^2}{|{\bf r_2}-{\bf r}'| |{\bf r_2}-{\bf r}''|} d\,^3 r' d\,^3 r'' d\,t' d\,t'' - 2 \int\int \frac{|\psi({\bf r}',t')|^2 |\psi({\bf r}'',t'')|^2}{|{\bf r_1}-{\bf r}'| |{\bf r_2}-{\bf r}''|} d\,^3 r' d\,^3 r'' d\,t' d\,t''\Big]\\
& + \frac{G^2m^4}{4\hbar^2}\Big[\int \frac{|\psi({\bf r}',t')|^2}{|{\bf r_1}-{\bf r}'|^2} d\,^3 r' d\,t' d\,t'' + \int \frac{|\psi({\bf r}',t')|^2}{|{\bf r_2}-{\bf r}'|^2} d\,^3 r' d\,t' d\,t''\\
& -2 \int \frac{|\psi({\bf r}',t')|^2}{|{\bf r_1}-{\bf r}'| |{\bf r_2}-{\bf r}'|} d\,^3 r' d\,t' d\,t'' \Big]
\label{phasesq}
\end{split}
\end{equation}
where the time integration is done up to a time $T$ and the volume integrations are over all of space.

To calculate the phase variance, we use the free gaussian solution of the wave function for a point particle of mass $m$
\cite{giulini2011gravitationally}  which is given by [we recall from our remark above (\ref{vtot}) that we are finding the  gravitational effect perturbatively] 
\begin{equation}
\psi(r,t) = (\pi a^2)^{-3/4} \Big(1+\frac{i\hbar t}{ma^2}\Big)^{-3/2}\exp\Big(-\frac{r^2}{2a^2(1+\frac{i\hbar t}{ma^2})}\Big)
\label{gaussian}
\end{equation}
Substitution of this free wave function in the above variance formula (\ref{phasesq}) gives the result
\begin{equation}
\Delta \phi^2 = \frac{G^2 m^4}{\hbar^2} T^2 \Big(\frac{1}{r_1} - \frac{1}{r_2}\Big)^2
\label{bigres}
\end{equation}
Details of this derivation are given in Appendix II. 
We demand the phase variance to be of the order $\pi^2 \sim 1$. This gives,
\begin{equation}
T \sim \frac{\hbar}{Gm^2} \cdot \frac{1}{|\frac{1}{r_1}-\frac{1}{r_2}|}
\end{equation} 
Hence
\begin{equation}
 T = \frac{\hbar}{|\Delta E|}; \qquad\qquad \Delta E = \frac{Gm^2}{r_1} - \frac{Gm^2}{r_2}
 \end{equation}
  Thus, from the stochastic Schr\"{o}dinger-Newton  equation we have deduced a result similar to the Di\'osi-Penrose criterion, namely that the decoherence time is of the order of the inverse of the gravitational potential energy difference at the two locations. This strongly suggests that gravitational decoherence drives the system to one of the solutions of the SN equation, as originally proposed by Di\'osi and Penrose. It is interesting that this result has been derived starting from the Einstein-Langevin equation, and it highlights the significance of considering quantum fluctuations in the stress tensor, while also encouraging us to believe that stochastic gravity is intimately connected with the stochastic SN equation.

Here, it is evident that we certainly cannot talk of the coherence length as simply the distance  $|r_1 - r_2|$ between the two points because the gravitational potential energy difference is involved. So, unlike the other gravity induced decoherence models, where the decoherence time and coherence length depend upon the distance between two points, here they actually depend upon the difference $|\frac{1}{r_1}-\frac{1}{r_2}|$. So for this case, we introduce a length scale $a_c$, which is the coherence length, defined as:
$$ \frac{1}{a_c} = \Big|\frac{1}{r_1}-\frac{1}{r_2}\Big|$$
Then from the above calculation we can write, 
\begin{equation}
T = \frac{\hbar a_c}{G m^2}
\end{equation}
Again we know that $T \sim m a_c^2/ \hbar$. These two expressions together give,
\begin{equation}
a_c = \frac{\hbar^2}{Gm^3}
\end{equation}
which is a well-known result. 

In our calculation, we have found that the decoherence time between states at the two points ${\bf r}_1$ and ${\bf r}_2$ depends on the difference between the gravitational potential energy  at these two points. This is somewhat similar, although not exactly same as, the so called Di\'osi - Penrose criterion. The latter says that if we have two stationary states represented by two mass configurations then their superposition will decay to one of the solutions of the SN equation in a time scale $\tau \approx \hbar / \Delta E_g$ where $\Delta E_g$ is the gravitational self-energy of the difference between the mass distributions of the two states \cite{Moroz:98, Penrose:96}. Penrose suggests in \cite{Penrose:96} that if we have two mass distributions representing the two states characterised by the mass densities $\varrho$ and $\varrho'$ respectively, then the measure of incompatibility between them would be \begin{equation}
\Delta = -4\pi G \int\int (\varrho({\bf x})-\varrho'({\bf x}))(\varrho({\bf y})-\varrho'({\bf y}))/|{\bf x}-{\bf y}| \,d^3{\bf x} \; d^3{\bf y}
\end{equation}
He suggests that the decay time would be of the form $\tau \sim \hbar / \Delta E_g$ where $\Delta E_g$ is $\Delta$ or some multiple of that quantity. Di\'osi also obtains the same form, as has been given in \cite{Diosi:89, Diosi:05}, where he finds that the decay time should be $\tau \sim \hbar / \Delta E_g$, with $\Delta E_g$ having a form
\begin{equation}
\Delta E_g = G \int\int [f({\bf x}|X)-f({\bf x}|X')][f({\bf x'}|X)-f({\bf x'}|X')]/|{\bf x}-{\bf x'}| \,d^3{\bf x}\; d^3{\bf x'}
\end{equation}
Here $f({\bf x}|X)$ denotes the mass density at ${\bf x}$ for a configuration denoted by $X$.

\section{Some qualitative estimates from the SN equation}
In an effort to obtain some estimates as to how the coherence length $a_c$ depends on the mass $m$ and size $R$ in case of an extended object, and in order to compare with the results of Karolyhazy, and of Di\'osi, we make some qualitative estimates.

Let us recall first the case of the point particle treated in \cite{giulini2011gravitationally} according to the SN equation:
\begin{eqnarray}
i\hbar\frac{\partial\psi({\bf r},t)}{\partial t} &=&\hat{H}_o\psi({\bf r},t)-Gm^2\Big(\int\frac{|\psi({\bf r}^{\,\prime},t)|^2}{|{\bf r}-{\bf r}^{\,\prime}|}d{\bf r}^{\,\prime}\Big)\psi({\bf r},t).
\end{eqnarray}
An estimate of the critical length and decay time can  be obtained from this equation considering the self-gravitation of an expanding wave packet. 
If we start with a spherically symmetric Gaussian wave packet of width $a$,
 \begin{equation}
 \psi(r,0)=(\pi a^{2})^{-3/4} \exp\left(-\frac{r^{2}}{2a^{2}}\right)
 \end{equation}
 then after time $t$, the wave function evolves, via Schr\"{o}dinger equation (in the absence of any gravitational potential), as the state given by (\ref{gaussian}).
 The radial probability density is at a maximum at $ r_{p} =a\sqrt{1+\frac{\hbar^{2}t^{2}}{m^{2}a^{4}}}$.
 So, the peak shifts with time with an acceleration given by the following equation:
 $
 \ddot{r}_{p}={\hbar^{2}}/{m^{2}r_{p}^{3}}
 .$
 This gives the acceleration of the wave packet when it expands freely.
 Now, the acceleration due to gravity for a point mass $m$ at a distance $r_{p}$ is 
 $
 \ddot{r}_{p}= {Gm}/{r_{p}^{2}}
 .$
 Let, after a certain time, these two accelerations become equal. We call the width of the wave packet in such an equilibrium as $a_{equil}$.
 Then equating the two accelerations we get, for a mass $m$,
 $
 m=({\hbar^{2}}/{Ga_{equil}}\Big)^{1/3}
 .$
 So, for a given mass, one can calculate this equilibrium width. If $r_{p}<a_{equil}$ then usual quantum evolution dominates while for $r_{p}>a_{equil}$ gravity becomes more significant and collapse of the wave function takes place. Hence, $a_{equil}$ is the same as as the critical length $a_c$.
 
 We now focus our attention on the behaviour of the wave function at $t=0$.
 At $t=0$ we have $r_{p}=a$.
 Now if we start with a mass $m$ such that ${\hbar^{2}}/{m^{2}a^{3}}>\frac{Gm}{a^{2}}$ then the Scr\"{o}dinger-like expansion keeps on accelerating unless the two effects become equal. In that case we have $a_{c}>a$. But if $\frac{\hbar^{2}}{m^{2}a^{3}}<\frac{Gm}{a^{2}}$ then the wave packet starts contracting right from the beginning and we get $a_{c}<a$. This can be summarized as below:
\begin{enumerate}
\item If $m<\Big(\frac{\hbar^{2}}{Ga}\Big)^{1/3}$ then $a_{c}>a$
\item If $m>\Big(\frac{\hbar^{2}}{Ga}\Big)^{1/3}$ then $a_{c}<a$
\end{enumerate} 
So, for an already contracted wave function, we must have $m>\Big(\frac{\hbar^{2}}{Ga}\Big)^{1/3}$.
$m=\Big(\frac{\hbar^{2}}{Ga}\Big)^{1/3}$ is the threshold mass for collapse.\\

\textbf{Critical Length for Extended Objects} :

We have already seen that the critical length for a single point-like mass $m$ is given by
$
a_c={\hbar^{2}}/{Gm^3}
$
Now we calculate the same for an extended object of mass $m$ and size $R$.
We consider two cases: (i)
$a_{c}\gg R$.
In this case, the gravitational acceleration on the wave packet which extends outside the mass $m$ is,
$
g={Gm}/{r_p^2}
$.
For the critical length $a_c$ we have,
$
{\hbar^{2}}/{m^2 a_c^3}={Gm}/{a_c^2}
$.
This essentially gives the same critical length as in case of a single point mass
$
a_c={\hbar^{2}}/{Gm^3}
$.
(ii) 
$a_{c}\ll R$.
When $a_{c}\ll R$, the wave packet lies inside the extended mass $m$. Considering the mass $m$ having uniform density within the radius $R$ we get the gravitational acceleration on the wave packet as,
$
g={Gmr_p}/{R^3}
$.
Again at the critical length,
$
{\hbar^{2}}/{m^2 a_c^3}={Gma_c}/{R^3}
$.
This gives the critical length as:
$
a_c={\hbar}/{Gm^{3}})^{1/4}R^{3/4}
$.
Interestingly, this same expression for $a_c$ is obtained by Di\'osi in his treatment of gravitational decoherence, while it differs from the expression obtained by Karolyhazy. What this is telling about possible similarity between the stochastic SN equation and the Di\'osi approach is not clear to us at present.

Putting $a_c \approx R$ gives the transition between micro region and macro region:
\begin{itemize}
\item micro region($m^3 R\ll\frac{\hbar^2}{G}$): critical length $a_c=\frac{\hbar^{2}}{Gm^3}$
\item macro region($m^3 R\gg\frac{\hbar^2}{G}$): critical length $a_c= \Big(\frac{\hbar^2}{Gm^{3}}\Big)^{1/4}R^{3/4}$
\item Transition occurs at $R\approx \frac{\hbar^2}{Gm^{3}}$
\end{itemize} 
 At $r_p=a_c$ the two opposite accelerations become equal and so, the wave function evolves at a constant rate. Let us consider a mass for which $m$ is below the threshold value i.e, $m<\Big(\frac{\hbar^2}{Ga}\Big)^{1/3}$. Initially when $r_p<a_c$, the wave packet expands at an accelerating rate until it reaches $r_p=a_c$. At this point, as the two forces become equal, it stops accelerating and starts expanding at a constant rate. As soon as $r_p$ becomes greater than $a_c$, again, gravity dominates and the wave packet shrinks to size $r_p=a_c$. Again at this point onwards, it keeps on contracting at a constant rate until $r_p<a_c$ where again Schr\"{o}dinger like expansion takes over. Thus, there should be an oscillation around the equilibrium width $a_c$ which has been found through numerical simulations in \cite{giulini2011gravitationally}. 

\bigskip

\bigskip


\section{Concluding remarks}
We believe that the stochastic SN equation we have found makes a useful contribution to the subject of gravity induced decoherence. The equation is possibly robust, because its origin lies in the well-defined theory of stochastic semiclassical gravity.  It correctly predicts the Di\'osi-Penrose decoherence criterion which relates decoherence time to the gravitational potential energy difference between the two points under consideration. The equation is on the same footing as the stochastic Schr\"{o}dinger equations of Karolyhazy and Di\'osi, especially in its linearised form obtained after using the approximate stochastic potential (\ref{vtot}), which choice makes the equation linear. The key difference of our equation is that it deals with the stochastic effects of self-gravity, whereas the Karolyhazy / Di\'osi models deal with stochastic effects induced by intrinsic space-time uncertainty. How these two effects relate to each other, and whether a more general treatment should include both effects remains a subject of future investigation. Like the master equations in the CSL and Di\'osi models, which resemble collisional decoherence models \cite{Vacchini:07}, our master equation in principle does the same, although explicit details remain to be worked out. 

Of course the full correlation function (\ref{fullcorr}) in our case  is state dependent, unlike in the case of the stochastic potentials introduced by Karolyhazy and by Di\'osi. Furthermore, the noise is not white, unlike in the case of Di\'osi's model, where the noise is white in time. And yet, the final physical results match in the two cases - the reasons for this intriguing feature are not clear to us at present.

In an interesting  recent paper Nimmrichter and Hornberger \cite{Nimmrichter:2015} have also proposed stochastic extensions of the (regularised) SN equation. One of their goals is to make the SN equation compatible with the no-signalling requirement. They suggest two possible stochastic extensions: one via discrete jumps described as a Poisson random process, and another as a continuous white noise (Wiener process). It is shown that nonlinearity, and hence superluminality, is avoided at the level of the stochastic average described by the density operator, which obeys a linear master equation. Our work differs from theirs in the key aspect that we derive the stochastic noise starting from the Einstein-Langevin equation, and this noise is different from the ones they propose. However, as noted above, if we work with the approximate potential (\ref{vtot}) [gravitational back-reaction is calculated iteratively] we also obtain a linear master equation, consistent with no signalling. Also, the noise that we introduce seems to lead naturally to the gravitational decoherence effect described by Di\'osi and Penrose.

The $N$-particle stochastic SN equation can be constructed following the method described in 
\cite{Bahrami:2014} (their Eqns. 6-12) and by using for the potential the form (\ref{stochasticeq}) given above. Generalizing the decoherence result obtained above, we expect this equation to produce localisation of macroscopic masses. Thus the two-particle stochastic SN equation can be expected to localise the two particles and correctly reproduce their classical Newtonian gravitational interaction.

An aspect which we have not yet investigated is the magnitude of stochastic heating brought about by the stochastic terms present in the new equation. It is known in the Di\'osi model that there are divergences in the master equation in the point particle limit, resulting in a divergent heating rate in this limit. One possible way to avoid this divergence is to introduce a cut-off (see for instance the detailed recent discussion in \cite{Bahrami:2014}). It is perhaps possible that this divergence problem does not arise in our model because the non-white noise in the non-Markovian equation induces a finite correlation length; however at the moment this is a conjecture, which needs to be scrutinised.

Further investigations should include studying the stochastic SN equation for states other than Gaussian wave-packets, and for extended objects. It will be useful also to see if there is some way to convert the stochastic equation into a collapse equation, and to compare it with CSL. One should also attempt to work out how the stochastic SN equation can be tested in experiments such as molecular interferometry and via gravitationally induced random diffusion \cite{Collett:2003,Bera2015}.

\bigskip

\noindent{\bf Acknowledgements:} RM thanks the staff of TIFR, where this work was done, for its kind hospitality. We would like to thank Mohammad Bahrami, Angelo Bassi, Lajos Di\'osi, Sandro Donadi, Thomas Durt, Bei-Lok Hu and Hendrik Ulbricht for useful discussions.  This work is supported by a grant from the John Templeton Foundation ({\#39530}).

S. B. and R. M. contributed equally to this work.

\bigskip


\bigskip

\centerline{\bf Appendix I : Master equation for the density matrix} 

We have from Eqn. (\ref{stochasticsnvector})
\begin{equation}
\frac{d|\Psi(t)\rangle}{d t} =\left[-\frac{i}{\hbar}\hat{H}_0 +i\sqrt{\gamma}\int d{\bf k}\langle \hat{A}^\dagger({\bf k})\rangle\hat{A}({\bf k}) + i\sqrt{\gamma}\int d{\bf k} \tilde{\xi}({\bf k},t)\hat{A}({\bf k})\right] |\Psi(t)\rangle\label{stochasticsnvector2}.
\end{equation}
It implies that
\begin{equation}
d|\Psi\rangle = -\frac{i}{\hbar}\hat{H}_0|\Psi\rangle dt+i\sqrt{\gamma}\int d{\bf k}\langle \hat{A}^\dagger({\bf k})\rangle\hat{A}({\bf k})|\Psi\rangle dt + i\sqrt{\gamma}\int d{\bf k} \tilde{\xi}({\bf k},t)\hat{A}({\bf k})|\Psi(t)\rangle dt,\label{diffstatevect}
\end{equation}
and, on taking the conjugate of the equation
\begin{equation}
d\langle\Psi| = dt\langle\Psi|\frac{i}{\hbar}\hat{H}_0 -dt\langle\Psi|i\sqrt{\gamma}\int d{\bf k}\langle \hat{A}({\bf k})\rangle\hat{A}^\dagger({\bf k})-dt\langle\Psi|i\sqrt{\gamma}\int d{\bf k} \tilde{\xi}^*({\bf k},t)\hat{A}^\dagger({\bf k}).\label{diffstatevectconj}
\end{equation}

The stochastic SN equation preserves the norm of the statevector $|\Psi\rangle$. To see this we compute the differential of the stochastic quantity $\langle\Psi|\Psi\rangle$:
\begin{equation}
d\langle\Psi|\Psi\rangle = d\langle\Psi||\Psi\rangle + \langle\Psi|d|\Psi\rangle
\end{equation}
From equations (\ref{diffstatevect}) and (\ref{diffstatevectconj})  we have
\begin{equation}
\begin{split}
d\langle\Psi|\Psi\rangle &= dt\langle\Psi|\frac{i}{\hbar}\hat{H}_0|\Psi\rangle -dt\langle\Psi|i\sqrt{\gamma}\int d{\bf k}\langle \hat{A}\rangle\hat{A}^\dagger|\Psi\rangle-dt\langle\Psi|i\sqrt{\gamma}\int d{\bf k} \tilde{\xi}\hat{A}^\dagger|\Psi\rangle\\
&-\langle\Psi|\frac{i}{\hbar}\hat{H}_0|\Psi\rangle dt+i\sqrt{\gamma}\langle\Psi|\int d{\bf k}\langle \hat{A}^\dagger\rangle\hat{A}|\Psi\rangle dt + i\sqrt{\gamma}\langle\Psi|\int d{\bf k} \tilde{\xi}^*\hat{A}|\Psi\rangle dt\\
&= i\sqrt{\gamma}\int d{\bf k}\left( \tilde{\xi}\langle\Psi|\hat{A}|\Psi\rangle-\tilde{\xi}^*\langle\Psi|\hat{A}^\dagger|\Psi\rangle\right) dt\\
&=0
\end{split}
\end{equation}

Let $\hat{\rho}_{st} = |\Psi\rangle\langle\Psi|$. Then
\begin{equation}
\begin{split}
d\hat{\rho}_{st} &= d|\Psi\rangle\langle\Psi|+|\Psi\rangle d\langle\Psi|\\
&=-\frac{i}{\hbar}[\hat{H}_0,\hat{\rho}_{st}]dt+i\sqrt{\gamma}\int d{\bf k}\left(\langle\hat{A}^\dagger\rangle\hat{A}\hat{\rho}_{st}-\hat{\rho}_{st}\hat{A}^\dagger\langle\hat{A}
\rangle\right)dt\\&+i\sqrt{\gamma}\int d{\bf k}\left(\hat{A}\tilde{\xi}\hat{\rho}_{st}-\hat{\rho}_{st}\tilde{\xi}^*\hat{A}^\dagger\right)dt.\label{densitymatsto}
\end{split}
\end{equation}

The density matrix of the system is the stochastic average of the $\hat{\rho}_{st}$
\begin{equation}
\hat{\rho} = \langle\hat{\rho}_{st}\rangle_s.
\end{equation}
We now take the stochastic average of Eqn. (\ref{densitymatsto}) and retain terms upto the order of $\sqrt{\gamma}$. {It simply means that $\langle\hat{A}\rangle \approx \langle\Psi_0|\hat{A}|\Psi_0\rangle $ where we have the expansion $|\Psi\rangle = |\Psi_0\rangle + \sqrt{\gamma}|\psi_1\rangle + \gamma|\Psi_2\rangle +\ldots$}
\begin{equation}
\begin{split}
\left\langle d\hat{\rho}_{st}\right\rangle_s &= \bigg\langle-\frac{i}{\hbar}[\hat{H}_0,\hat{\rho}_{st}]dt+i\sqrt{\gamma}\int d{\bf k}\left(\langle\hat{A}^\dagger\rangle\hat{A}\hat{\rho}_{st}-\hat{\rho}_{st}\hat{A}^\dagger\langle\hat{A}
\rangle\right)dt\\&+i\sqrt{\gamma}\int d{\bf k}\left(\hat{A}\tilde{\xi}\hat{\rho}_{st}-\hat{\rho}_{st}\tilde{\xi}^*\hat{A}^\dagger\right)dt\bigg\rangle_s\\
\frac{d\hat{\rho}}{dt}&=-\frac{i}{\hbar}[\hat{H},\hat{\rho}] +i\sqrt{\gamma}\int d{\bf k}\left(\langle\hat{A}^\dagger\rangle\hat{A}\hat{\rho}-\hat{\rho}\hat{A}^\dagger\langle\hat{A}\rangle\right)+i\sqrt{\gamma}\int d{\bf k}\left(\hat{A}\left\langle\tilde{\xi}\hat{\rho}_{st}\right\rangle_s-\left\langle\hat{\rho}_{st}\tilde{\xi}^*
\right\rangle_s\hat{A}^\dagger\right).\label{densitymatrixequation}
\end{split}
\end{equation}
According to the Furutsuâ-€"Novikov formula
\begin{equation}
\begin{split}
\left\langle F[\tilde{\xi}]\tilde{\xi}({\bf k},t)\right\rangle_s =\int d{\bf k}^{\,\prime}\int dt^\prime D({\bf k},{\bf k}^{\,\prime},t,t^\prime)\left\langle\frac{\delta F[\tilde{\xi}]}{\delta \tilde{\xi}({\bf k}^{\,\prime},t^\prime)}\right\rangle_s,
\end{split}
\end{equation}
where $F[\tilde{\xi}]$ is an arbitrary functional of the gaussian stochastic noise and $D(k,k^{\,\prime},t,s)$ is 
the correlation function of $\xi({\bf k},t)$
\begin{equation}
\begin{split}
D(k,k^{\,\prime},t,t^\prime) &= \langle\xi({\bf k},t),\xi({\bf k}^{\,\prime},t^\prime)\rangle_c\\
&=\left\langle kk^\prime\int d{\bf r}_1d{\bf r}_2e^{-i{\bf k}.{\bf r}_1}e^{-i{\bf k}^{\,\prime}.{\bf r}_2}\int d{\bf r}^{\,\prime}d{\bf r}^{\,\prime\prime} \frac{\xi_{00}({\bf r}^{\,\prime},t)}{|{\bf r}_1-{\bf r}^{\,\prime}|}\frac{\xi_{00}({\bf r}^{\,\prime\prime},t^\prime)}{|{\bf r}_2-{\bf r}^{\,\prime\prime}|}\right\rangle_c\\
&=  kk^\prime\int d{\bf r}_1d{\bf r}_2e^{-i{\bf k}.{\bf r}_1}e^{-i{\bf k}^{\,\prime}.{\bf r}_2}\int d{\bf r}^{\,\prime}d{\bf r}^{\,\prime\prime} \frac{N_{0000}({\bf r}^{\,\prime},{\bf r}^{\,\prime\prime},t,t^\prime)}{|{\bf r}_1-{\bf r}^{\,\prime}||{\bf r}_2-{\bf r}^{\,\prime\prime}|},\\
\end{split}
\end{equation}
where $N_{0000}({\bf r}^{\,\prime},{\bf r}^{\,\prime\prime},t,t^\prime)$ is given by Eqn. (\ref{2pointcorre}).
 
Let $F[\tilde{\xi}]= \hat{\rho}_{st}$. We assume a system with spherical symmetry. Then,
\begin{equation}
\begin{split}
\left\langle\hat{\rho}_{st}\tilde{\xi}(k,t)\right\rangle_s = 4\pi\int dk^{\,\prime}k^{\prime 2}\int dt^\prime D(k,k^{\,\prime},t,t^\prime)\left\langle\frac{\delta \hat{\rho}_{st}}{\delta \tilde{\xi}(k^\prime,t^\prime)}\right\rangle_s.\label{computedensity}
\end{split}
\end{equation}

We now transit to the interaction picture where the operators and state vectors are defined as
\begin{equation}
\hat{A}({\bf k},t) = e^{i\frac{\hat{H}_0t}{\hbar}}\hat{A}({\bf k})e^{-i\frac{\hat{H}_0t}{\hbar}}, \qquad |\Psi (t)\rangle_I = e^{i\frac{\hat{H}_0t}{\hbar}}|\Psi (t)\rangle.
\end{equation}
We, then, expand $|\Psi\rangle_I$ with the parameter $\sqrt{\gamma}$ as follows
\begin{equation}
|\Psi\rangle_I = |\Psi_0\rangle_I + \sqrt{\gamma}|\psi_1\rangle_I + \gamma|\Psi_2\rangle_I +\ldots\label{intpicexpan}
\end{equation}

In the interaction picture Eqn. (\ref{stochasticsnvector2}) takes the form
\begin{equation}
\frac{d|\Psi(t)\rangle_I}{d t} =i\sqrt{\gamma}\int d{\bf k}\left(\langle \hat{A}^\dagger({\bf k})\rangle + \tilde{\xi}({\bf k},t)\right)\hat{A}({\bf k},t) |\Psi(t)\rangle_I\label{stochasticsnvector2int}.
\end{equation}

On plugging Eqn. (\ref{intpicexpan}) in Eqn. (\ref{stochasticsnvector2int}) and collecting the terms of order $0$, $\sqrt{\gamma}$ and $\gamma$ we get
\begin{itemize}
\item[$\mathcal{O}(0):$]\begin{equation}
\frac{\partial}{\partial t}|\Psi_0\rangle_I = 0
\end{equation}
\item[$\mathcal{O}(\sqrt{\gamma}):$]
\begin{equation}
\begin{split}
\frac{\partial}{\partial t}|\Psi_1\rangle_I &= i\int d{\bf k}\left(\langle \hat{A}^\dagger({\bf k})\rangle + \tilde{\xi}({\bf k},t)\right)\hat{A}({\bf k},t) |\Psi_0\rangle_I\\
|\Psi_1(t)\rangle &= i\int_0^tdt^\prime\int d{\bf k}\left(\langle \hat{A}^\dagger({\bf k})\rangle + \tilde{\xi}({\bf k},s)\right)\hat{A}({\bf k},t^\prime-t)|\Psi_0(t)\rangle\label{gammaorder1}
\end{split} 
\end{equation}
\item [$\mathcal{O}(\gamma):$] and so on\ldots
\end{itemize}

The density matrix in Schr\"{o}dinger picture $\hat{\rho}_{st} = |\Psi\rangle\langle\Psi|$, when expanded perturbatively upto order $\sqrt{\gamma}$ is 
\begin{equation}
\hat{\rho}_{st} = |\Psi_0\rangle\langle\Psi_0|+\sqrt{\gamma}(|\Psi_1\rangle\langle\Psi_0|+|\Psi_0\rangle\langle\Psi_1|).
\end{equation}
Using Eqn. (\ref{gammaorder1}), we can evaluate 
\begin{equation}
\frac{\delta\hat{\rho}_{st}}{\delta \tilde{\xi}({\bf k}^{\,\prime},t^\prime)} = i\sqrt{\gamma}(\hat{A}({\bf k}^{\,\prime},t^\prime-t)\hat{\rho}_{st}^0),
\end{equation}
and
\begin{equation}
\frac{\delta\hat{\rho}_{st}}{\delta \tilde{\xi}^*({\bf k}^{\,\prime},t^\prime)} =-i\sqrt{\gamma}(\hat{\rho}_{st}^0\hat{A}^\dagger({\bf k}^{\,\prime},t^\prime-t)),
\end{equation}
where $\hat{\rho}_{st}^0 = |\Psi_0\rangle\langle\Psi_0|$.

Hence from Eqn. (\ref{computedensity}), we have
\begin{equation}
\begin{split}
\left\langle\hat{\rho}_{st}\tilde{\xi}(k,t)\right\rangle_s &= 4\pi i\sqrt{\gamma}\int dk^{\,\prime}k^{\prime 2}\int_0^t dt^\prime D(k,k^{\,\prime},t,t^\prime)\left\langle\hat{A}({\bf k}^{\,\prime},t^\prime-t)\hat{\rho}_{st}^0\right\rangle_s\\
&= 4\pi i\sqrt{\gamma}\int dk^{\,\prime}k^{\prime 2}\int_0^t dt^\prime D(k,k^{\,\prime},t,t^\prime)(\hat{A}({\bf k}^{\,\prime},t^\prime-t)\hat{\rho})
\end{split}.
\end{equation}

Finally from Eqn. (\ref{densitymatrixequation}) we obtain
\begin{equation}
\begin{split}
\frac{d\hat{\rho}}{dt} &=-\frac{i}{\hbar}[\hat{H},\hat{\rho}] +i\sqrt{\gamma}\int d{\bf k}\left(\langle\hat{A}^\dagger\rangle\hat{A}\hat{\rho}-\hat{\rho}\hat{A}^\dagger\langle\hat{A}\rangle\right)+i\sqrt{\gamma}\int d{\bf k}\left(\hat{A}\left\langle\tilde{\xi}\hat{\rho}_{st}\right\rangle_s-\left\langle\hat{\rho}_{st}\tilde{\xi}
\right\rangle_s\hat{A}^\dagger\right)\\
&= -\frac{i}{\hbar}[\hat{H},\hat{\rho}] +i\sqrt{\gamma}\int d{\bf k}\left(\langle\hat{A}^\dagger\rangle\hat{A}\hat{\rho}-\hat{\rho}\hat{A}^\dagger\langle\hat{A}\rangle\right)\\&-
\gamma\int_0^t dt^\prime \int dkdk^\prime \tilde{D}(k,k^{\,\prime},t,t^\prime)\bigg(\hat{A}(k)\hat{A}(k^\prime,t^\prime-t)\hat{\rho}+\hat{\rho}\hat{A}^\dagger(k^\prime,t^\prime-t)\hat{A}^\dagger(k)\bigg),
\end{split}
\end{equation}
where $\tilde{D}(k,k^{\,\prime},t,t^\prime) = 16\pi^2k^2k^{\prime 2}D(k,k^{\,\prime},t,t^\prime)$. We note that the calculation has been carried out perturbatively, by expanding the state in powers of
$\sqrt{\gamma}$, and is exact up to order  $\sqrt{\gamma}$.

\bigskip

\centerline{\bf Appendix II: Calculation of phase variance for gaussian state}

Let us consider the first term in (\ref{phasesq}). The integration can be separated into two integrals of $r'$ and $r''$. The integration can be performed in spherical polar co-ordinates. Similarly, all the other terms can also be calculated. Some steps of the calculations are as follows.
The first term can be separated into two integrals of $r'$ and $r''$ of the form $\int\int \frac{|\psi({\bf r}',t')|^2}{|{\bf r_1}-{\bf r}'|} d\,^3 r' d\,t'$.
We have, 
\begin{equation}
|\psi({\bf r}',t')|^2 = \pi^{-3/2} \frac{1}{a^3 (1 + \frac{\hbar^2 t'^2}{m^2 a^4})^{3/2}} \exp\Big(-\frac{r'^2}{a^2 (1 + \frac{\hbar^2 t'^2}{m^2 a^4})}\Big)
\end{equation}
Now let ${\bf r}_1-{\bf r}'={\bf r}$ so that $r'^2=r_1^2 +r^2 -2r_1 r \cos \theta$. This gives 
\begin{equation}
 |\psi({\bf r}',t')|^2 = \pi^{-3/2} \frac{1}{a^3} \alpha^{3/2} \exp\Big(-\frac{\alpha}{a^2}(r_1^2 +r^2 -2r_1 r \cos \theta)\Big)
 \end{equation}
  where $\alpha = 1/\big(1+\frac{\hbar^2 t'^2}{m^2 a^4}\big)$.
 
The integration is now straightforward. Other terms can also be calculated in the same fashion.\\
Small time approximation: In our calculation, in order to obtain an analytical result, the time integration has been calculated using small time approximation i.e. the interval $T$ is such that
\begin{equation}
e^{-r_1^2/a^2} \frac{\hbar^2 r_1}{m^2 a^4}\frac{T^2}{3a \sqrt{\pi}} \ll {\rm erf}(r_1/a)
\label{erf}
\end{equation}
where the error function ${\rm erf}(x)$ is given, along with the imaginary error function ${\rm erfi}(x)$, used below, as
\begin{equation}
{\rm erf}(x) = \frac{2}{\sqrt \pi} \int _0^x e^{-t^{2}} \ dt \ ;\qquad {\rm erfi}(x) = -i{\rm erf}(ix) =\frac{2}{\sqrt \pi} \int _0^x e^{t^{2}}\ dt
\end{equation}
Th condition given in (\ref{erf}) is well satisfied for $r_1/a \to \infty$.\\
After using small time approximation we get
\begin{equation}
\begin{split}
\Delta \phi^2 &= \frac{3G^2m^4}{4\hbar^2}T^2\left(\frac{1}{r_1^2}\left({\rm erf}\left(\frac{r_1}{a}\right)\right)^2+\frac{1}{r_2^2}\left({\rm erf}\left(\frac{r_2}{a}\right)\right)^2-\frac{2}{r_1r_2}{\rm erf}\left(\frac{r_1}{a}\right){\rm erf}\left(\frac{r_2}{a}\right)\right)\\
&+ \frac{G^2m^4}{4\hbar^2}T^2\bigg(\frac{\sqrt{\pi}}{r_1a}e^{-\frac{r_1^2}{a^2}}{\rm erfi}\left(\frac{r_1}{a}\right)+\frac{\sqrt{\pi}}{r_2a}e^{-\frac{r_2^2}{a^2}}{\rm erfi}\left(\frac{r_2}{a}\right)\\
&-\frac{2 \sqrt{\pi}}{a\sqrt{r_1 r_2}}e^{-1/2(r_1^2/a^2 + r_2^2/a^2)}\sqrt{{\rm erfi}(r_1/a){\rm erfi}(r_2/a)}\bigg)\\
\end{split}
\label{phadiff}
\end{equation}

The last term in the above equation has not been calculated explicitly as it was very complicated. So we used the argument that since it was initially obtained by breaking a squared term, the final result should be a perfect square. Putting the last term in this form gives the desired result.
%

We now calculate term by term in Eqn. (\ref{phadiff}) in the limit $r_1/a \to \infty$ and $r_2/a \to \infty$.
First we look at the terms involving $r_1$. So, the first and fourth terms of the above expression together give,
\begin{equation}
\frac{G^2 m^4}{4 \hbar^2}\frac{T^2}{r_1^2}\Big[\sqrt{\pi}\frac{r_1}{a}e^{-r_1^2/a^2}{\rm erfi}\Big(\frac{r_1}{a}\Big) + 3 {\rm erf}^2\Big(\frac{r_1}{a}\Big)\Big]
\end{equation}
As $\frac{r_1}{a} \to \infty$, the term inside [..] goes to $4$. So, we have, this term as
\begin{equation}
\frac{G^2 m^4}{4 \hbar^2}\frac{T^2}{r_1^2} \cdot 4
\end{equation} 
Similarly, by adding second and fifth terms, we get,
\begin{equation}
\frac{G^2 m^4}{4 \hbar^2}\frac{T^2}{r_2^2} \cdot 4
\end{equation}
Now let us consider the third and sixth terms together:
\begin{equation}
-\frac{G^2 m^4}{4 \hbar^2} \frac{2T^2}{r_1 r_2}\Big[3 {\rm erf}\Big(\frac{r_1}{a}\Big){\rm erf}\Big(\frac{r_2}{a}\Big) +\sqrt{\frac{\pi r_1 r_2}{a^2}} e^{-\frac{1}{2}(r_1^2/a^2 + r_2^2/a^2)} \sqrt{{\rm 
erfi}\Big(\frac{r_1}{a}\Big){\rm erfi}\Big(\frac{r_2}{a}\Big)}\Big]
\end{equation}
Again the term in square brackets goes to 4 for $\frac{r_1}{a} \to \infty$ and $\frac{r_2}{a} \to \infty$. We get,
\begin{equation}
-\frac{G^2 m^4}{4 \hbar^2} \frac{2T^2}{r_1 r_2} \cdot 4
\end{equation}
Finally, adding all the terms gives the desired result
\begin{equation}
\Delta \phi^2=\frac{G^2 m^4}{\hbar^2} T^2 \Big(\frac{1}{r_1} - \frac{1}{r_2}\Big)^2
\end{equation}
which is Eqn. (\ref{bigres}).

The calculations can be repeated for the case $\frac{r_1}{a} \approx 1$ and $\frac{r_2}{a} \approx 1$ and the results come out to be the same. The only problem in this limit seems to be that the small time approximation is not valid for microscopic masses. For large masses, it is still valid.

\bigskip


\centerline{\bf REFERENCES}

\bibliography{biblioqmts3}

\end{document}